\documentclass[twocolumn,prd,aps,showpacs]{revtex4}

\def\be{\begin{equation}}
\def\ee{\end{equation}}
\def\ba{\begin{eqnarray}}
\def\ea{\end{eqnarray}}

\def\rate{{\cal R}}
\def\ampl{{\cal A}}
\def\const{{\rm const.}~}
\def\Ecr{E_{\rm cr}}
\def\uplus{a_{\uparrow +}}
\def\uminus{a_{\uparrow -}}
\def\dplus{a_{\downarrow +}}
\def\dminus{a_{\downarrow -}}
\def\Psiu{\Psi_{\uparrow}}
\def\Psid{\Psi_{\downarrow}}
\def\x{{\bf x}}
\def\y{{\bf y}}
\begin{document}
\title{Quasiparticle scattering by quantum phase slips in one-dimensional superfluids}
\author{S. Khlebnikov}
\affiliation{Department of Physics, Purdue University, West Lafayette, 
IN 47907, USA}
\begin{abstract}
Quantum phase slips (QPS) in narrow superfluid channels generate momentum by
unwinding the supercurrent. In a uniform Bose gas, this momentum needs to be 
absorbed by quasiparticles (phonons). 
We show that this requirement results in an
additional exponential suppression of the QPS rate (compared to the rate of QPS
induced by a sharply localized perturbation). In BCS-paired fluids, momentum
can be transferred to fermionic quasiparticles, and we find an interesting
interplay between quasiparticle scattering on QPS and on disorder. 
\end{abstract}
\maketitle
{\em 1. Introduction.}
Particle production and scattering during tunneling are of interest in many 
applications. They can significantly modify the transition amplitude and, 
in addition, destroy
coherence between the states connected by tunneling. Of particular interest
are situations when some scattering {\em must} occur, as a consequence
of a nonzero momentum generated by the tunneling event.

One of the simplest such cases, and the one on which we concentrate in this 
Letter, is a quantum phase slip in a narrow superfluid channel. Advances in
containment and cooling of atomic Bose gases
make it possible to contemplate a ring (toroidal) geometry---a
channel closed in the longitudinal ($x$) direction.
This is the case considered here, although many of our results are applicable 
also to a cigar-shaped geometry. In addition, since these gases are dilute, the
correlation (``healing'') length $\xi$ can be made larger than the transverse size,
making the system effectively one-dimensional (1D).

In a 1D system, the order parameter $\psi$ can momentarily vanish at some
point, allowing the phase of $\psi$ to unwind. Such events are referred to 
as phase slips.
Here we consider the case of very low temperatures when the main mechanism
for phase slips is quantum, rather than thermal, fluctuations. These are
quantum phase slips (QPS). Their high-temperature counterpart, thermally activated
phase slips (TAPS), was considered in the framework of the nucleation theory of
Refs. \cite{Little,LA,McCH} in Ref. \cite{TAPS}.

The purpose of this Letter is to present a calculation of how the QPS rate is
affected by interactions of QPS with quasiparticles (phonons).
First, we show that unless there is a highly localized (or very strong) external
perturbation, QPS without momentum transfer to phonons are strongly suppressed by
destructive interference. This makes such QPS highly unlikely and makes 
interaction with phonons a principal effect.
Next, we compute the rate of QPS assisted by phonon scattering and find that 
at low temperature $T$ it is given by $\rate \sim \exp(-v_s P/2T)$, 
where $P = 2\pi n$ is the momentum
generated by unwinding the phase; $n$ is the linear number density of the gas, 
$v_s$ is the speed of Bogoliubov's phonons. We also estimate the preexponent.
The physical interpretation
of this result is that the destructive interference is removed when momentum
$P$ is absorbed by the phonon subsystem, but that requires energy of at least
$v_s P / 2$ (to allow for tunneling between phonon states with total 
momenta $-P/2$ and $P/2$).

The above estimate for the rate is the main result of the present paper. 
It applies to a uniform weakly-coupled Bose gas at temperatures $T \ll gn$
and represents a strong suppression, compared to the case when QPS is induced 
by a highly localized perturbation, and momentum conservation plays no role.
We briefly discuss means to experimentally detect phonon scattering on
QPS in atomic superfluids. 
Finally, we compare this case to the case of BCS paired superconductors.

{\em 2. QPS without momentum transfer to phonons.}
By unwinding the supercurrent, each QPS releases momentum $\pm P$.
(In the ring-shaped geometry, a better conserved
quantity would be the angular momentum, but this 
distinction will not be important in what follows.) This corresponds to an 
imaginary contribution to the Euclidean action of a QPS:
\be
\Delta S_{\rm QPS} = - 2\pi i \int_0^{x_0} n(x) dx \approx - 2\pi i \bar{n} x_0 \; ,
\label{DS}
\ee
where $x_0$ is the QPS location, and $\bar{n}$ is an average density. 
Then, the tunneling amplitude is 
\be
\ampl = \int dx_0 {\cal B}(x_0) e^{2 \pi i \bar{n} x_0} \; .
\label{ampl}
\ee
If the scale of variation of ${\cal B}(x_0)$ (due to possible 
variations of density, etc.) is some large $\ell$, 
the amplitude (\ref{ampl}) is suppressed by $\exp(-\const \bar{n} \ell)$.
For $\ell \gg \xi$, this suppression is much stronger than the semiclassical
$\exp(-\const v_s/g)$. It can be seen as a consequence of the destructive 
interference among QPS with different values of $x_0$. 

One way to avoid this suppression is to consider a localized perturbation
with scale $\ell \ll \xi$. Calculations for this case have been done in Ref.
\cite{Kagan&al}.
Another possibility, which we consider here, is that the momentum released by
a QPS is absorbed by phonons. We expect that if phonons change their total momentum
by $2\pi \bar{n}$, the recoil on the QPS will be zero, and the suppression by 
$\exp(-\const \bar{n} \ell)$ will be lifted. For sufficiently uniform gases, such
phonon-assisted tunneling may be the dominant QPS mechanism. 
Furthermore, the two mechanisms can be distinguished 
experimentally by measuring the momentum transfer (see below).

{\em 3. Phonon production at $T=0$}. 
In this case, the phonon subsystem is initially in the ground state, so to 
transfer momentum to phonons we need to produce them in the final state. To set
up come notation, we briefly discuss this case and show why, if the only source
of energy available for phonon production is the energy due to unwinding of
the supercurrent, such processes are kinematically forbidden. The formalism
sketched here may be useful when there are other sources of energy, 
as for instance in various ``cosmology in the lab'' proposals \cite{cosmo}, 
where the properties of the superfluid are changed in time to mimic certain
cosmological spacetimes.

We consider a weakly-coupled 1D Bose gas.
In 1D, the effective coupling constant $g$ has dimension of velocity, and the
dimensionless measure of the coupling strength is 
\be
\frac{g}{v_s} = 0.008
\left( \frac{a}{10 {\rm nm}} \right)^{1/2}
\left( \frac{10^{14} {\rm cm}^{-3}}{n^{(3)}} \right)^{1/2}
\left( \frac{1 \mu{\rm m}}{R} \right)^2 ,
\label{meas}
\ee
where $v_s$ in the phonon speed, $a$ is the scattering length,
$n^{(3)}$ is the 3D number density, 
and $R$ is the transverse radius of the channel.
As we will see, the real part of the instanton action is 
of order $\pi v_s / g$ (times a logarithm).

The crossover between the weakly and strongly coupled regimes can be located 
using the
results of Haldane \cite{Haldane}: it occurs at $g/ v_s \approx \pi$.
In what follows we assume a somewhat weaker coupling
$g / v_s \sim 1$, which according to (\ref{meas}) can be achieved for
$R \sim 0.1~\mu$m. Note that such small transverse radii may be necessary in any
case, just to drive the system into the 1D regime. Indeed, the 1D regime 
requires $R \ll \xi$, while the typical values of $\xi$ are a few tenths of micron.

So, if it were not for the destructive interference, an instanton action of order
$S_{\rm QPS}\sim \pi v_s / g$ would allow for a relatively large QPS rate in narrow 
channels.

Since the system is in the weakly coupled regime, we can obtain a more precise estimate
for $S_{\rm QPS}$ by using the Gross-Pitaevsky (GP) description. This is given by
the Euclidean Lagrangian
\be
L_E =  \psi^{\dagger} \partial_\tau \psi + \frac{1}{2M} |\partial_x \psi|^2 
+ {g \over 2} |\psi|^4 - \mu |\psi|^2 \; .
\label{L1}
\ee
where $\tau= it$ is the Euclidean time.
Outside the QPS core, fluctuations of density are small, and we can linearize the
system by writing $\psi =\sqrt{n + \delta n} \exp(i\theta)$ and expanding in
small $\delta n$. Integrating out $\delta n$ and assuming the long-wavelength limit, 
we obtain
\be
L_E \approx i n \partial_\tau \theta +{1\over 2g} (\partial_\tau \theta)^2
+ {n \over 2 M} (\partial_x \theta)^2 \; .
\label{L2}
\ee
From this, we read off the sound velocity $v_s = \sqrt{g n / M}$.
In what follows, we assume that $n$ is $x$-independent and choose our unit of length 
so that $v_s = 1$.

Since (\ref{L2}) is a long-wavelength limit, it implies a certain choice of 
the reference frame:
momenta of produced phonons should be small in comparison with $1/\xi$. Thus, the
Galilean invariance of the original theory (\ref{L1}) now holds only approximately
and only for boost speeds much smaller than $(M\xi)^{-1} \sim v_s$. 
This approximate symmetry, however, would be completely sufficient to restore 
the topological (first) term in eq. (\ref{L2}).

Instantons corresponding to QPS are vortices of the theory (\ref{L1}) in the 
$(x,\tau)$ plane. Using the limit (\ref{L2}), one finds that 
away from the core the phase $\theta$ of the instanton solution is
\be
\theta_I(x-x_0, \tau-\tau_0) = \arg[x-x_0 +i(\tau-\tau_0)] \; ,
\label{inst}
\ee
where $(x_0,\tau_0)$ is the location of the QPS. 

Consider an instanton-antiinstanton pair in the presence of a persistent supercurrent.
The corresponding classical configuration is
\be
\theta_c(x,\tau) = \theta_I(x-x_0, \tau-\tau_0) - \theta_I(x,\tau) 
+ \frac{2\pi}{L} N x 
\label{theta}
\ee
(here we have placed the antiinstanton at the origin). The last term corresponds to
a supercurrent with winding number $N$ over the longitudinal length $L$ of the
system. Substituting (\ref{theta}) into 
(\ref{L2}) and integrating, we obtain the Euclidean action of the
pair to logarithmic accuracy
\be
S_{\rm pair} = 
-2\pi i n x_0 - E \tau_0 + 
\frac{2\pi}{g} \ln\frac{d}{\xi} + O\left( \frac{1}{g} \right) 
\; .
\label{S3}
\ee
where $d = (x_0^2 + \tau_0^2)^{1/2}$, and 
$\xi = (4 g M n)^{-1/2}$ is the healing length (the size of the vortex core).
Eq. (\ref{S3}) applies when the separation $d$ is much smaller than the 
longitudinal size $L$ of the channel. Contribution from the QPS core is subleading
and included in the $O(1/g)$ term.

The first term in (\ref{S3}) is the imaginary part anticipated in (\ref{DS}).
The second term, with 
\be
E= E_N \equiv \frac{(2\pi)^2 N}{gL} \; ,
\label{E}
\ee
comes from the cross term between the
instanton and the supercurrent contributions to $\partial_x \theta$. 
Note that $E_N$
is precisely the energy released by a single phase slip. 

It is easy to see, however, that this energy alone is not sufficient
to satisfy the kinematic condition for production of phonons with total
momentum $P=2\pi n$. Indeed, the minimal
(critical) energy required for that is
\be
\Ecr = 2\pi n = \frac{\pi} {g\xi} \; .
\label{Ecr}
\ee
Landau's criterion for superfluidity is $N/L < 1 / 4\pi \xi$, so in the superfluid 
state $E_N < \Ecr$ ($E_N = \Ecr$ precisely at the critical current).
We conclude that at zero temperature
QPS are effectively impossible, unless one considers short-scale perturbations
or introduces other sources of energy. This conclusion
is consistent with Galilean invariance (GI) of the 
$T=0$ state. One way to break GI is to include thermal fluctuations.

{\em 4. Phonon scattering at $T\neq 0$}. 
In this case, there are preexisting phonons in the system, and we consider
a QPS assisted by scattering of phonons, so that the momentum released by the 
QPS is absorbed by the phonon subsystem. We take the limit when the
energy $E_N$ is negligible compared to the energy of the requisite
phonon
state. At $T\neq 0$, the system can be regarded as compactified on a cylinder of 
circumference $\beta = 1/T$. Equivalently, instead of a single IA pair we should
consider a periodic chain of such pairs:
\be
\theta_c = \arg[1 - e^{-i(\tau-\tau_0) -(x-x_0)} ] - \arg(1 - e^{-i\tau -x}) \; .
\label{chain}
\ee
Here, instantons are shifted relative
to antiinstantons by amount $x_0$ in the $x$-direction, and by $\tau_0$ in the 
$\tau$-direction. These are ``soft'' collective coordinates, which will need to
be integrated over.
All lengths in (\ref{chain}) are in units of $\beta / 2\pi$,
and we assume that the total longitudinal length $L$ is effectively infinite.

The action per period, to logarithmic accuracy, equals
\be
S_{\rm chain} = -i P x_0 + \frac{2\pi}{g} \ln\frac{1}{\xi} 
+ \frac{\pi}{g} \ln ( \cosh x_0 - \cos \tau_0 ) \; .
\label{S4}
\ee
The last term is the instanton-antiinstanton interaction. 
If it were not for the momentum $P$, the dominant configuration would have
$x_0 = 0$ and coincide with the periodic instanton \cite{periodic}, which
describes tunneling between quasiparticle states with zero total momenta.
At nonzero $P$, configurations with $x_0\neq 0$ become important.

The integrals over $x_0$ and $t_0 = -i\tau_0$ can be done by steepest descent.
In the limit $\beta \gg \xi$ (that is, $T \ll gn$), the saddle point is at 
$\tau_0 = \pi$ and $x_0 \approx i\pi$. The
saddle-point action is dominated by the first term in (\ref{S4}):
$S_{\rm chain} \approx P/ 2T$ (where we have returned to the physical units of length).
Corrections to this result are controlled by the small parameter $T\xi$.

The above action determines the exponential factor in the QPS rate. 
The preexponent is given by the determinant of field fluctuations near
the saddle point. As usual, the main effect is due to the collective coordinates.
Two zero modes, corresponding to translations in space and time, each contribute
a factor of order $1/\xi\sqrt{g}$, up to a power of $\ln(\beta/\xi)$. 
The saddle-point integrals over 
$x_0$ and $\tau_0$ result each in a factor
of order $(k_* \xi)^{-1/2}$, where $k_*$ is the typical momentum of phonons
scattered by the QPS. Other modes result in a factor of order unity.
Assembling all the factors together, we obtain, for the QPS rate
per length $L$, in the limit $\beta \gg \xi$,
\be
\rate \sim \frac{v_s^2 L}{g \xi^3 k_*} e^{-v_s P/2T} 
\label{R}
\ee
(where we have restored $v_s$).
A detailed study of the initial and final phonon states (to be presented elsewhere)
gives $k_*^3 \sim (\xi^2 \beta)^{-1} \ln(\beta / \xi)$, in the same limit. So,
$k_* \ll \xi^{-1}$, which confirms that the
longwave limit (\ref{L2}) is applicable to the present problem, with corrections
controlled by $k_* \xi$.

The exponential in  (\ref{R}) can be interpreted by noting that at a finite 
temperature the quasiparticle system 
can tunnel from a state with momentum $-P/2$ to a state with momentum $P/2$,
thus absorbing the full momentum $P$ but requiring energy of only $P/2$.
The main physical effect reflected in (\ref{R}) is that transfer of a
large momentum requires tunneling from a fairly high-energy state, resulting
in an additional suppression. If no momentum transfer were necessary, the leading
term in the exponent would be $(2\pi/g) \ln (T\xi)$ \cite{periodic} (see also
Ref. \cite{Kagan&al}), so the suppression would be much weaker.

Phonon scattering by QPS in atomic superfluids can in principle be detected 
experimentally by measuring the momentum distribution of atoms, using, say, the
standard momentum imaging. In the ring-shaped
geometry, this can be done by opening the trap in one place and 
looking at the atomic cloud emitted from there. For example, if we start with 
the state without supercurrent, a single QPS will create a unit of supercurrent 
and a compensating normal flow in the opposite direction. Since, the
``normal'' atoms have comparatively large momenta, the cloud will be highly asymmetric.

{\em 5. Quasiparticle production in superconductors.}
The case of BCS superconductors is especially interesting because nonzero resistance
has been observed 
in ultra-thin wires at low temperatures \cite{exp1,exp2}, and it
has been interpreted in terms of QPS \cite{thin2}.

Curiously, the ``hydrodynamic'' part of the effective Lagrangian for superconductors
is given by the same expression (\ref{L1}) as for atomic superfluids, except that
$\psi$ is now the field of Cooper pairs, which describes charge fluctuations
in the wire (cf. Ref. \cite{LE}). Accordingly, 
the coupling constant is $g=4e^2/C$, where $C$ is the capacitance per unit 
length. The reason for this similarity is
that in 1D screening is weak, and the plasmon mode is gapless 
(the Mooij-Sch\"{o}n mode \cite{MS}). From (\ref{L1}), we obtain the speed
of plasmons as $c_0 = (e^2 n_s / m_* C)^{1/2}$,
where $n_s \equiv 4 \langle \psi^\dagger \psi \rangle m_*/M$, and $m_*$ is the
electron mass. This is indeed the speed of the Mooij-Sch\"{o}n mode.

For BCS superconductors, there is an additional Lagrangian describing 
the interaction of $\psi$ with fermionic quasiparticles. 
Indeed, production of these quasiparticles is the least energy-consuming 
way
of momentum production, since each quasiparticle carries a large momentum $k_F$
but costs only a relatively small energy equal to the superconducting gap $\Delta$.
This mechanism is unavailable to atomic superfluids.

On general grounds, one expects that impurities, which break both the Galilean
invariance and momentum conservation, may significantly affect the physics of QPS
\cite{eff_action}. Here we present an explicit example of such interplay,
using the simple, if
somewhat artificial, single-channel limit, in which electrons are 
confined to motion in the $x$ direction only, and their mean-free path 
$\lambda$ is taken to be much larger than the QPS core size.

The Fermi surface now consists
of two points, at $k=\pm k_F$, so there are right-movers, denoted by
subscript $(+)$, and left-movers, with subscript $(-)$. Each variety
can have up or down
spin, corresponding to $\sigma= \uparrow, \downarrow$. 
We define new fermionic operators
$a_{\sigma\pm}$ related to the original electron operators $c_{\sigma\pm}$ by
\be
c_{\sigma\pm}(x,t) =e^{\pm ik_F x} a_{\sigma \pm} (x,t) \; .
\label{orig}
\ee
The new operators can be conveniently assembled into two Dirac spinors:
$\Psiu = (\uplus, \uminus)$, $\Psid = (\dplus^{\dagger}, \dminus^{\dagger})$.

In the case without disorder, $\lambda \to \infty$,
the Lagrangian of fermions can be written in the relativistic form:
\be
L_F = i\sum_{\sigma=\uparrow,\downarrow}
{\overline \Psi}_\sigma \gamma^{\mu} \partial_{\mu} \Psi_\sigma -
g' \psi {\overline \Psiu} \Psid - g' \psi^* {\overline \Psid} \Psiu \; .
\label{L22}
\ee
Here $\mu=0,1$, the (1+1) $\gamma$-matrices are $\gamma^0=\sigma_1$,
$\gamma^1=-i\sigma_2$, and ${\overline \Psi} = \Psi^{\dagger} \gamma_0$.
In this section, we set $v_F=1$. 

For any $\psi$ with a nontrivial winding at infinity, the Euclidean
equation of motion for $\Psi$ has a nontrivial normalizable solution 
(zero mode). The zero mode is known explicitly for $\psi$ of the form
$\psi(r,\vartheta) = f(r)\exp(i\vartheta)$ (in polar coordinates) \cite{JR}, 
and its existence
in the general case is guaranteed by an index theorem \cite{Weinberg}.
The zero mode causes the amplitude for tunneling without fermions to vanish
identically. As has been explained
by 't Hooft \cite{tHooft} in the context of gauge theories in four dimensions,
this means that only certain
anomalous Green functions, involving fermions, are nonzero. In 
our case, the relevant Green function is 
$\langle \uminus^\dagger(x) \uplus(y) \rangle$. It corresponds to creation of
a quasiparticle-quasihole pair at the QPS core. The total momentum 
produced in this process is $2k_F$. The phenomenon
is a 1D analog of ``momentogenesis'' \cite{momen} by vortex motion 
in a 2D array of Josephson junctions.

Note that $2k_F$ is precisely the momentum
necessary to cancel the momentum $-P$ released by unwinding of the current. 
Indeed, $P = 2\pi \langle \psi^\dagger \psi \rangle = \pi n$, where 
$n$ is the electron density. Since $n  = 2k_F /\pi$, we have $P = 2k_F$.
If the spatial dependence of the zero mode for $\uplus$ is $\chi(x-x_0)$, then
for the original operator $c_{\uparrow +}$ it is $\exp[ik_F(x-x_0)] \chi(x-x_0)$.
In this way, we find
\ba
\lefteqn{\langle c^\dagger_{\uparrow -}(\x) c_{\uparrow +}(\y) \rangle \sim}  
& & 
\nonumber \\
& & 
\int dx_0 d\tau_0 e^{-S'} e^{ik_F(x + y)} 
\chi(\x - \x_0) \chi(\y - \x_0) \; ,
\label{GF}
\ea
where $\x=(x,\tau)$, and $S'$ is the real part of the instanton action.

At $T=0$ and in the absence of disorder, production of a quasiparticle-quasihole
pair via the matrix element (\ref{GF}) is kinematically forbidden in the same way
as phonon production was in the case of an atomic superfluid (Landau's criterion
now being that the velocity of the superflow must be smaller than the depairing velocity
$\Delta / k_F$). However, scattering on an impurity potential $W(x)$
introduces an additional Lagrangian
\be
\Delta L_F = W(x) \sum_{\sigma} 
c^\dagger_{\sigma -} c_{\sigma +} + {\rm H.c.} \; .
\label{loc}
\ee
Computing to the first order in $W$ and absorbing the fermion operators into
the anomalous correlator (\ref{GF}), we find that in the presence of disorder
the amplitude of QPS without quasiparticle production becomes nonzero.

I thank A. J. Leggett, Y. Lyanda-Geller, and M. Shaposhnikov for 
discussions, and the Institute of Theoretical Physics at University of 
Lausanne, where part of this work was done, for hospitality.
This work was supported 
in part by the U.S. Department of Energy through Grant DE-FG02-91ER40681 
(Task B).


\begin{thebibliography}{99}
\bibitem{Little} W. A. Little, Phys. Rev. {\bf 156}, 396 (1967).
\bibitem{LA} J. Langer and V. Ambegaokar, Phys. Rev. {\bf 164}, 498 (1967).
\bibitem{McCH} D. McCumber and B. Halperin, Phys. Rev. B {\bf 1}, 1054 (1970).
\bibitem{TAPS} E. J. Mueller, P. M. Goldbart, and Y. Lyanda-Geller,
Phys. Rev. A {\bf 57}, R1505 (1998).
\bibitem{Kagan&al} Yu. Kagan, N. V. Prokofiev, and B. V. Svistunov,
Phys. Rev. A {\bf 61}, 045601 (2000).
\bibitem{cosmo} C. Barcel\'{o}, S. Liberati, and M. Visser, gr-qc/0305061,
and references therein.
\bibitem{Haldane} F. D. M. Haldane, Phys. Rev. Lett. {\bf 47}, 1840 (1981).
\bibitem{periodic} S. Khlebnikov, V. Rubakov, and P. Tinyakov, 
Nucl. Phys. {\bf B367}, 334 (1991).
\bibitem{exp1} N. Giordano, Phys. Rev. Lett. {\bf 61}, 2137 (1988); Physica B
{\bf 203}, 460 (1994).
\bibitem{exp2} A. Bezryadin, C. N. Lau, and M. Tinkham, 
Nature (London) {\bf 404}, 971 (2000); C. N. Lau {\em et al.}, 
Phys. Rev. Lett. {\bf 87}, 217003 (2001).
\bibitem{thin2} A. D. Zaikin, D. S. Golubev, A. van Otterlo, 
and G. T. Zim\'{a}nyi, Phys. Rev. Lett. {\bf 78}, 1552 (1997);
D. S. Golubev and A. D. Zaikin, Phys. Rev. B {\bf 64}, 014504 (2001).
\bibitem{LE} A. Luther and V. J. Emery, Phys. Rev. Lett. {\bf 33}, 589 (1974).
\bibitem{MS} J. E. Mooij and G. Sch\"{o}n, Phys. Rev. Lett. {\bf 55}, 114 (1985).
\bibitem{eff_action} A. van Otterlo, D. S. Golubev, A. D. Zaikin, and G. Blatter,
Eur. Phys. J. B {\bf 10}, 131 (1999).
\bibitem{JR} R. Jackiw and P. Rossi, Nucl. Phys. B {\bf 190}, 681 (1981).
\bibitem{Weinberg} E. J. Weinberg, Phys. Rev. D {\bf 24}, 2669 (1981).
\bibitem{tHooft} G. 't Hooft, Phys. Rev. D {\bf 14}, 3432 (1976).
\bibitem{momen} Yu. G. Makhlin and G. E. Volovik, Pis'ma ZhETF {\bf 62}, 923
(1995) [JETP Lett. {\bf 62}, 941 (1995)].
\end{thebibliography}
\end{document}